\documentclass[a4paper]{jpconf}
\usepackage{graphicx}
\usepackage{amssymb}
\usepackage{amsmath}
\usepackage{relsize}
\usepackage{citesort}

\begin{document}
\title{Phenomenology of hadron structure --- why low energy physics matters}

\author{A.~Courtoy}

\address{IFPA, Inst. de Physique, Universit\'e de Li\`ege, Belgium \\
        INFN-LNF, Frascati, Italy}
\ead{aurore.courtoy@ulg.ac.be}

\begin{abstract}
The description of the internal structure of hadrons is one of the main goal of QCD. At moderate energy scales, the hadronic representation succeeds to the partonic description, rendering challenging the description of the dynamics of scattering processes and hadronic structure. The information on the hadron structure is embodied in the long distance contributions which are defined as Parton Distribution Functions (PDFs). PDFs are a key  framework for connecting the low and high-energy regimes, in that the knowledge on non-perturbative QCD  carries important consequences at the high-energy level. We here review recent progress in the description of the proton, from complementary approaches such as fits of PDFs, phenomenological analyses and experimental predictions in view of the JeffersonLab upgrade and applications for high-energy colliders.
\end{abstract}

\section{Introduction}
\vspace{.5cm}

QCD describes interactions as the exchange of a quantum degree of freedom called ``color".  The theory of  strong interactions has the interesting property of asymptotic freedom.  As the running coupling constant $\alpha_s(Q^2)$ becomes smaller, the perturbative treatment of QCD allows for the explanation of the hadronic phenomena, whose basic ingredients are the Parton Distributions.

On the other hand, hadrons that are actually observed in nature carry no color --- they are color singlets. So far, physicists have failed to describe colorless hadrons within QCD due to the property of confinement:
At low energy, there is no justification for  a perturbative treatment of QCD.  We do not longer have a description of the relevant low energy observables from QCD. In this regime, non-perturbative approaches  come into play. 

In these proceedings, we will describe how, in these schemes,  parameters are fixed by phenomenology. The two regimes of the strong interactions form an undivided whole, though the transition of the relevant degrees of freedom is not fully understood yet. As such, low energy parameters ---as accounting for the physical phenomena--- will guide  high-energy observables. Among the many interesting implications of non-perturbative physics for perturbative QCD and high-energy observables, we will consider 2 main directions. The first is related to the PDFs. The quark degrees of freedom transition into hadronic degrees of freedom at a scale intimately related to the initial scale chosen for the PDF fits.  We here discuss various approaches of the determination of the transition scale in Sections~\ref{sect:hadro_scale} $\&$ \ref{sect:NP_scale}. The first steps toward ``soft"  evolution are discussed in Section~\ref{sect:NP_scale}.  The phenomenological role of the running coupling constant in the infrared region is highlighted in Section~\ref{sect:soft_evo}.
This analysis induced a need to reconsider the PDFs at  large values of Bjorken-$x$, close to the exclusive limit. Interestingly, the large-$x$ PDF carry important consequences at the high-energy level as is shown in Section~\ref{sect:largex}.
The second direction we want to explore, in Section~\ref{sect:charges}, relates hadronic matrix elements to observables of physics Beyond the Standard Model. In particular, the scalar and tensor charges have been shown to be of great interest for precision measurements of new physics.

\section{Parton Distribution Functions at High Energy}
\vspace{.5cm}

     A way of connecting the perturbative and non-perturbative worlds has traditionally been through the study of Parton Distribution Functions (PDFs): 
    Deep Inelastic processes are such that they enable us  to look with a good resolution inside the hadron and allow us to resolve the very short distances, {\it i.e.} small configurations of quarks and gluons. This part of the process is described through perturbative QCD.  A resolution of such short distances is obtained with the help of non-strongly interacting probes.  Such a probe, typically a photon, is provided by hard reactions. In that scheme, the PDFs reflect how the target reacts to the probe, or how the quarks and gluons are distributed inside the target. The insight into the structure of hadrons is reached at that stage:  the large virtuality of the photon, $Q^2$, involved in such processes allows for the factorization of  the hard (perturbative) and soft (non-perturbative) contributions in their amplitudes ---in an Operator Product Expansion style.  They are matrix elements of a bilocal current on the light-cone,
\begin{eqnarray}
P^{\mu} q(x)&=&  \int \frac{d\tau}{4\pi}\, e^{ix\tau} \left \langle PS\right | {\bar q}(0) \gamma^{\mu} q(\tau n)\left |PS\right\rangle\quad,
\label{eq:pdf}
\end{eqnarray}
with $n^{\mu}=(1,0,0,-1)/(\sqrt{2} \Lambda)$ a light-like  vector. The distribution functions are non-perturbative objects as they describe the large distance behavior of hadrons, a regime where confinement starts to matter. 
The virtuality of the photon introduces  {\it  the factorization scale}, {\it i.e.} PDFs explicitly depend on $Q^2$.
This  $Q^2$-evolution is dictated by the Dokshitzer-Gribov-Lipatov-Altarelli-Parisi (DGLAP) equations,
 \begin{eqnarray}
Q^2\,\frac{\partial}{\partial Q^2} q(x, Q^2)&=& \frac{\alpha_s(Q^2)}{2\pi}\, \int_x^1 \frac{d\xi}{\xi}\, P\left(\frac{x}{\xi}, \alpha_s(Q^2)\right)\, q(\xi, Q^2)\quad,
\label{eq:DGLAP}
\end{eqnarray}
here at leading order in $\alpha_s$--- and where $P(x, Q^2)$ are the splitting functions.

At leading order ({\it leading-twist}), there are three types of
PDFs: the ordinary number density, the helicity, and the transversity. The former, $q(x)$, is the well-known unpolarized PDF and is accessible through, {\it e.g.}, inclusive deep inelastic scattering (DIS), {\it i.e.} in the Bjorken regime.
On the other hand,  $g_1(x)$, the helicity PDF, is less constrained. The experimental knowledge on  $h_1(x)$, the transversity PDF, is sparse as it is a chiral-odd quantity, not accessible through fully inclusive processes.

Collinear PDFs are universal and come into play in many processes, {\it e.g.} Drell-Yan processes, proton-proton collisions, $\cdots$ and are therefore of the utmost importance, especially for precision measurements.
 
Unpolarized PDFs have been extensively studied from first principles, in models and fits. The global fits combine  various data set with different energy range. The precision obtained in the various fits, for light flavor and valence quarks, is high ; gluon and sea quarks as well as low and large-$x$ regions can incontestably be improved. For example, in  DIS,  
the hard scattering part of the process
can be presently described using splitting functions up to next-to-next-to-leading order (NNLO) \cite{Moch:2004pa,Vogt:2004mw}, and Wilson coefficients functions up to N$^3$LO
\cite{Vermaseren:2005qc}. 
PDF parametrizations along with their uncertainties  have been obtained applying this framework up to NNLO, by a number of collaborations (see review in  Ref.~\cite{Forte:2013wc}). We cite MSTW, CTEQ, NNPDF, HERAFitter, GJR, ABM.\footnote{A useful link to compare PDF sets can be found in Ref.~\cite{durham} or the proceedings of the PDF4LHC working groups.}

The parameterizations present statistical error on the PDF itself as well as on the value of $\alpha_s(M_Z^2)$. The propagation of  these uncertainties for the LHC phenomenology has been extensively studied, see {\it e.g.} \cite{Forte:2013wc}. A representative example is the Higgs production through gluon-gluon fusion: $\sim 7\%$ of uncertainty on the cross section comes from uncertainty on PDF and $\alpha_s(M_Z^2)$. Those numbers vary from one parameterization to the other.\footnote{Fig.~7 of Ref.~\cite{Forte:2013wc} is very illustrative.}

In the standard PDF fitting approaches,  an {\it arbitrary} initial scale for the evolution equations, $Q_0^2>1$ GeV$^2$, is chosen: In parton distribution analyses the $x$-dependence of the PDFs is thus extracted at a particular scale $Q_0^2$, usually referred to as the {\it input scale}. A first systematic study of the effects of the choice of the input scale in global determinations of parton distributions and QCD parameters has been presented in Ref.~\cite{JimenezDelgado:2012zx}, introducing the concept of {\it procedural bias} in PDF analyses. 
The first consequence of the choice of the input scale in a parton distribution analysis is that it affects the determination  the strong coupling constant $\alpha_s(M_Z^2)$. In fact, the latter is determined  together with the parton distributions and is substantially correlated with the gluon distribution which drives the QCD evolution.
Besides, at low scales, there is an ambiguity related to the hadronic representation. The evolution equations modify the PDFs' radiative behavior in opposition to the low-energy valence behavior. The author of Ref.~\cite{JimenezDelgado:2012zx} uses a dynamical GJR parton distribution that optimally determines the input scale~\cite{Gluck:2007ck}. The latter is used as a guideline for the corresponding degrees of freedom ---for $Q_0^2<1$ GeV$^2$ the PDF should tend to a valence-like behavior.
$Q_0^2$ turns out to be of the order of $0.55$GeV$^2$ (with $\Lambda_{\mbox{\tiny NLO}}^{n_f=3}\sim303$MeV), {\it i.e.},  in the region where non-perturbative inputs cannot be neglected, in a top-down approach.

In the next Sections, we will see how this non-perturbative input can come into play.

\section{Hadronic Scale} 
\label{sect:hadro_scale}
\vspace{.5cm}

It is still a challenge to describe consistently the dynamics of scattering processes and hadronic structure at moderate energy scales. Because  at such a  moderate scale the hadronic representation gives way to the partonic description, it is called {\it the hadronic scale}. The hadronic scale is peculiar to each hadronic representation and should be related to the {\it input scale} defined in the previous Section.

\subsection{Determination of $Q_0$: Standard Approach}
\vspace{.5cm}

From a bottom-up point of view, the evaluation of PDFs  is guided by a standard scheme, set up in valuable litterature of the 90s \cite{Traini:1997jz,Stratmann:1993aw,Parisi:1976fz}. This scheme runs in 3 main steps.
    First, we either build models consistent with QCD in a moderate energy range, typically the hadronic scale; or we use effective theories of QCD for the description of hadrons at  the same  energy range. Second,  PDFs are evaluated in these models, giving a description of the Bjorken-$x$ dependence of the distribution. Third, the scale dependence of these distributions is studied. The last step allows to bring the moderate energy description of hadrons to the {\it factorization scale}, thanks to the QCD evolution equations~(\ref{eq:DGLAP}). Here we are interested in the  matching of  non-perturbative models to perturbative QCD, using experimental data.
    
 The hadronic scale is defined at a point where the partonic content of the model, defined through the second moment of the parton distribution,  is known. For instance, the CTEQ parameterization gives~\footnote{MSTW gives a similar result.}
 \begin{equation}
 \left \langle  (u_v+d_v)(Q^2=10 \mbox{GeV}^2)) \right\rangle_{n=2}=0.36\quad,
 \end{equation}
 with $q_v$ the valence quark distributions and with $\langle q_v (Q^2)\rangle_n=\int_0^1 dx \, x^{n-1} \, q_v(x, Q^2)$.
 Scenarios for the hadronic representation have to be chosen. In an extreme scenario,  {\it i.e.}, when we assume that the partons are pure valence quarks,  the second Mellin moment is evolved downward until
   \begin{equation}
 \left \langle  (u_v+d_v)(\mu_0^2) \right\rangle_{n=2}=1\quad.
 \end{equation}
 The hadronic scale is found to be $\mu_0^2 \sim 0.1$ GeV$^2$.

   This standard procedure to fix the hadronic (non-perturbative) scale  pushes perturbative QCD to its limit. 
   In effect, the hadronic scale turns out to be of a few hundred MeV$^2$, where the strong coupling constant has already started approaching its Landau pole. As it will be shown hereafter, the N$^m$LO evolution converges very fast, what justifies the perturbative approach.
   Consequently, the behaviour of the strong coupling constant plays a central role in the QCD evolution of parton densities. We here extend the standard procedure with the non-perturbative generalization of the QCD running coupling.


\subsection{$Q_0$ from Non-Perturbative Physics}
\vspace{.5cm}

We call perturbative evolution the renormalization group equations (RGE). 
The running of the coupling constant is driven by the RGE. In QCD, $\alpha_s$ is defined by renormalization conditions imposed at a large momentum scale where the coupling is small. The running coupling constant is dimensionless, but through dimensional transmutation, 
the strength of the interaction may be described by a dimensionful parameter.  QCD scale, $\Lambda_{\mbox{\tiny QCD}}$, is then defined as the energy scale where the interaction strength reaches the value $1$. 

\begin{figure}
\begin{center}
\includegraphics[scale= .6]{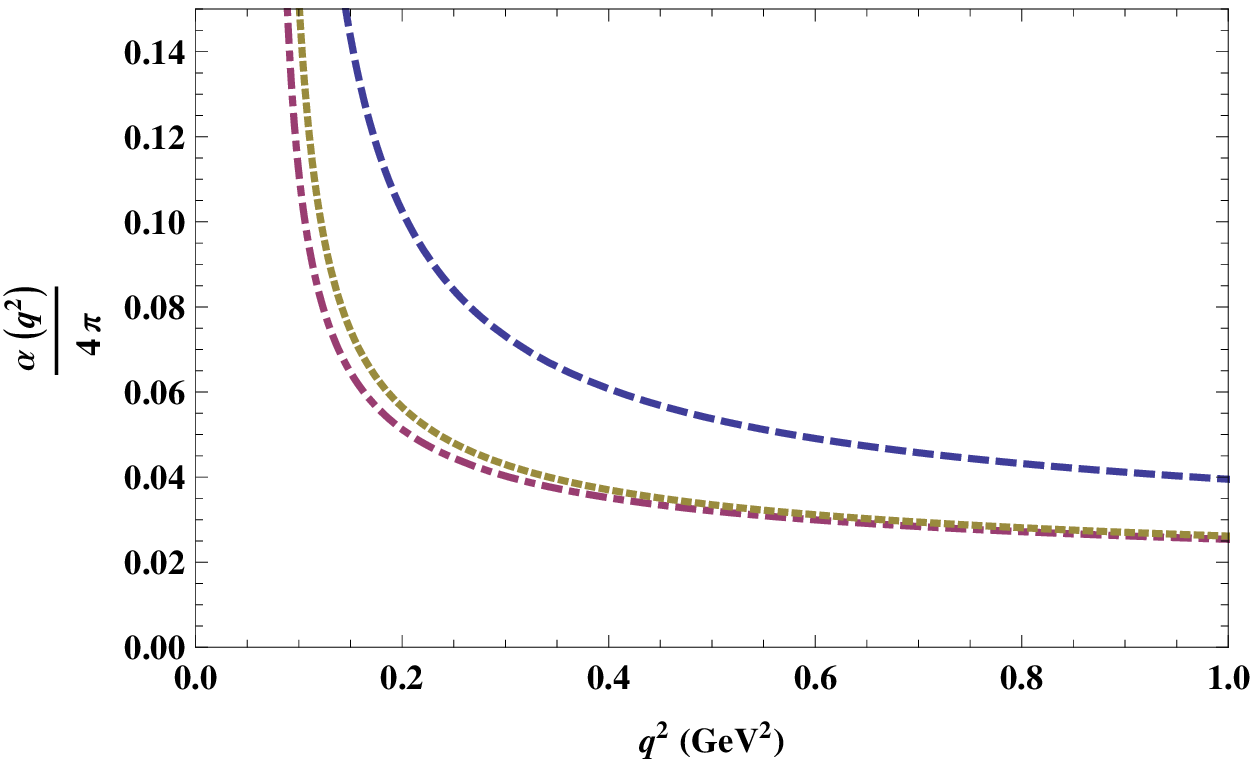}
\includegraphics[scale= .6]{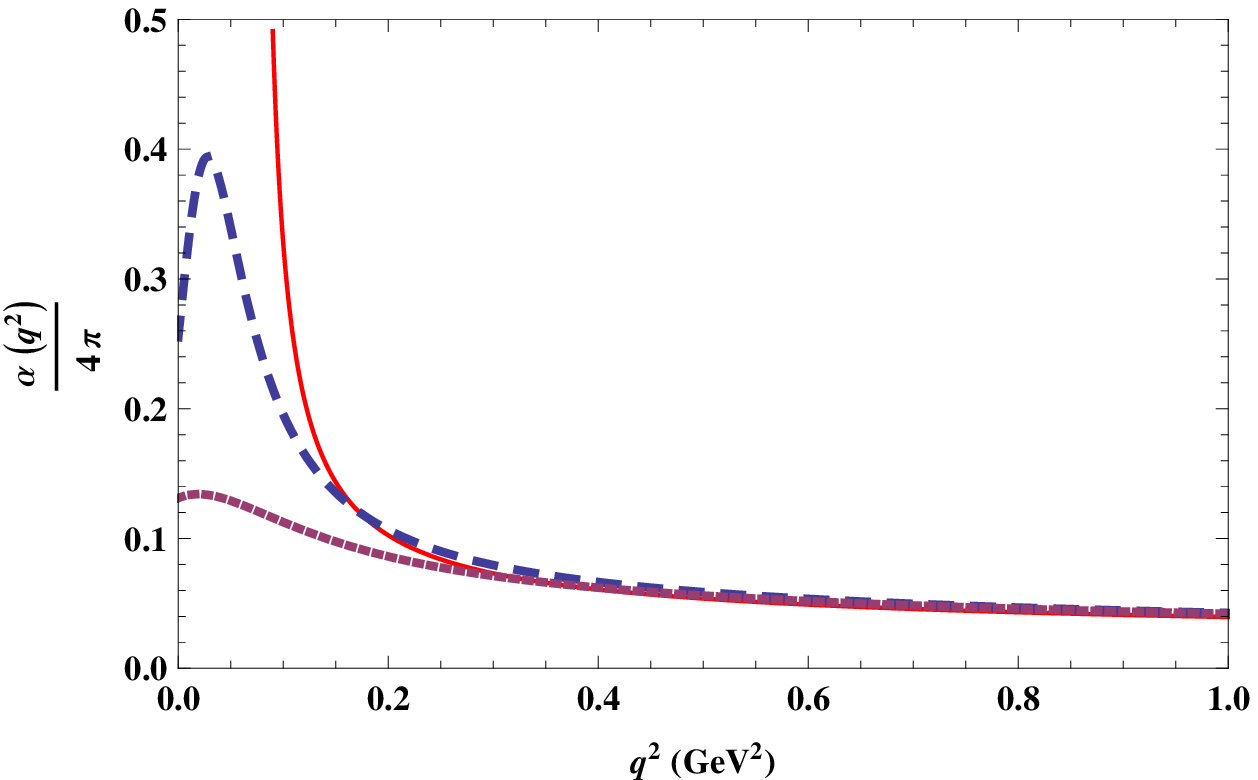}
\caption{ The running of the  coupling constant. {\it Left panel}: The short dashed curve  corresponds to the LO solution, the medium dotted-dashed curve to NLO solution and the tiny dashed curve to the NNLO solution ($\Lambda = 250$ MeV). {\it Right panel}: The running of the effective coupling. The dotted and dashed curves represent the non-perturbative evolution with the parameters $m_0=0.3$GeV and, respectively, for the medium dashed blue curve $\rho=1.53$, for the short dashed purple curve $\rho=2.2$. The solid curve shows the NLO evolution with $\Lambda = 250$ MeV.}
\label{aperb} 
\label{comparisons}
\end{center}
\end{figure}

At N$^m$LO the scale dependence of the coupling constant is given by

$$\frac{d \, a (Q^2)}{ d(\ln \;Q^2)}  = \beta_{\mbox{\tiny N}^m\mbox{\tiny LO}}(\alpha_s) =\stackrel{m}{\sum_{k=0}} a^{k+2} \beta_k,$$
where 
$a = {\alpha_s / 4 \pi}.$
We  show here the solution to $k=2$, {\it  i.e.}, NLO.\footnote{
$
\beta_0 =  11 - \frac{2}{3}\, n_f  \quad,
\beta_1 =  102 -\frac{38}{3} \,n_f $, 
where $n_f$ stands for the number of effectively massless quark flavors and $\beta_k$ denote the coefficients of the usual four-dimensional $\overline{MS}$ beta function of QCD.}
The evolution equations for the coupling constant can be integrated out exactly leading to
\begin{alignat}{2}
& \ln (Q^2/\Lambda_{\mbox{\tiny LO}}^2)&&=\frac{1}{\beta_0 a_{\mbox{\tiny LO}} } \quad ,\nonumber\\
& \ln (Q^2/\Lambda_{\mbox{\tiny NLO}}^2) &&= \; \frac{1}{\beta_0 a_{\mbox{\tiny NLO}}} +  \frac{b _1}{\beta_0} \ln (\beta_0 a_{\mbox{\tiny NLO}})- \frac{b _1}{\beta_0} \ln (1 + b_1 a_{\mbox{\tiny NLO}}) \quad ,
%
\label{exact}
\end{alignat}
%
 %
where $b_k = {\beta_k / \beta_0}$. These equations, except the first, do not admit closed form solution for the coupling constant, and we have solved them numerically. We show their solution, for the same value of $\Lambda = 250$ MeV, in  Fig. \ref{aperb}.

We see in Fig.~\ref{aperb} (left panel) that the NLO and NNLO solutions agree quite well even at very low values of $Q^2$.  They agree  even better if we change  the value of $\Lambda$ for the NNLO slightly, confirming the fast convergence of the expansion.  This analysis concludes, that even close to the Landau pole, the convergence of the perturbative expansion is quite rapid, specially if we use a different value  of $\Lambda$ to describe the different orders, a feature which comes out from the fitting procedures.
{\it This fast convergence ensures that perturbative evolution can still be used at rather low scales.} However, when entering the non-perturbative regime, other mechanisms take place that influence the QCD evolution. That is what we will call here non-perturbative evolution.

It is well established by now that the QCD running coupling (effective charge)  
freezes in the deep infrared. 
This non-perturbative property can be understood from various non-perturbative approaches~\cite{Cornwall:1981zr,Fischer:2003rp,Aguilar:2009nf,Shirkov:1997wi,Mattingly:1992ud}, {\it e.g.} from the point of view of
the dynamical gluon mass generation \cite{Cornwall:1981zr}.\footnote{Even though  the  gluon is massless  at the  level  of the  fundamental  QCD Lagrangian, and  remains
massless to all order in perturbation theory, the non-perturbative QCD
dynamics  generate  an  effective,  momentum-dependent  mass,  without
affecting    the   local    $SU(3)_c$   invariance,    which   remains
intact.}  
At the level of the Schwinger-Dyson equations
the  generation of such a  mass is associated with 
the existence of 
infrared finite solutions for the gluon propagator, 
i.e. solutions with  $\Delta^{-1}(0) > 0$.
Such solutions may  
be  fitted  by     ``massive''  propagators  of   the form 
$\Delta^{-1}(Q^2)  =  Q^2  +  m^2(Q^2)$;
$m^2(Q^2)$ is  not ``hard'', but depends non-trivially  on the momentum  transfer $Q^2$.
One physically motivated possibility, which we shall use in here, is  the so called logarithmic mass running, which is defined by
\begin{equation}
m^2 (Q^2)= m^2_0\left[\ln\left(\frac{Q^2 + \rho m_0^2}{\Lambda^2}\right)
\bigg/\ln\left(\frac{\rho m_0^2}{\Lambda^2}\right)\right]^{-1 -\gamma}.
\label{eq:mass}
\end{equation}

\begin{figure}
\begin{center}
\includegraphics[scale= .8]{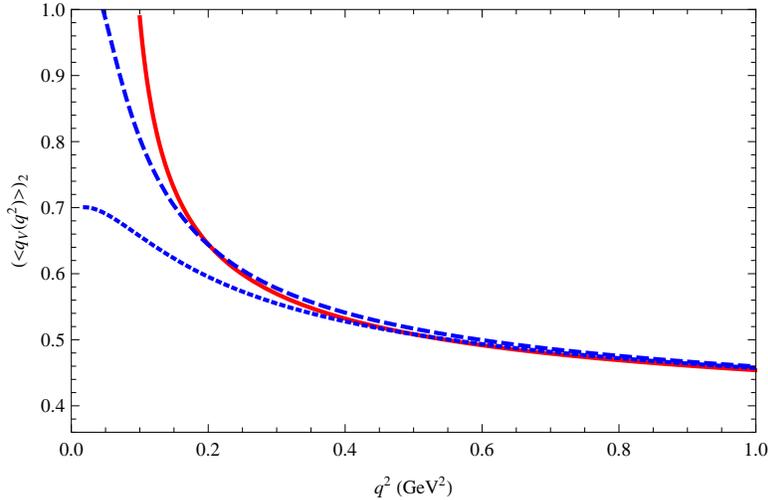}
\caption{ Left: The running of the effective coupling. The dotted and dashed curves represent the non-perturbative evolution with the parameters used above. The solid curve shows the NNLO evolution with $\Lambda = 250$ MeV.
Right: The evolution of the second moment of the valence quark distribution. The solid curve represents the perturbative LO approximation. }
\label{alpha}
\label{qval}
\end{center}
\end{figure}

Note that when $Q^2\to 0$ one has $m^2(0)=m^2_0$. Even though in principle we do not have 
any theoretical constraint that would put an upper bound to the value of $m_0$, 
phenomenological estimates place it in the range $m_0 \sim \Lambda - 2 \Lambda$~\cite{Bernard:1981pg,Parisi:1980jy}. The other parameters were fixed at $\rho \sim 1-4$,  ${\gamma} = 1/11$ \cite{Cornwall:1981zr,Aguilar:2007ie,Aguilar:2009nf}. 
%
 The  non-perturbative  generalization  of $\alpha_s(Q^2)$
the  QCD  running  coupling, comes in the form
\begin{equation}
a_{\mbox{\tiny NP}}(Q^2) = \left[\beta_0 \ln \left(\frac{Q^2 +\rho m^2(Q^2)}{\Lambda^2}\right)\right]^{-1} ,
\label{alphalog}
\end{equation}
where we use the same notation as before and NP stands for Non-Perturbative. Note that its zero gluon mass limit leads to the LO perturbative 
coupling constant momentum dependence.
The $m^2(Q^2)$ in the argument of the logarithm 
tames  the   Landau pole, and $a(Q^2)$ freezes 
at a  finite value in the IR, namely  
\mbox{$a^{-1}(0)= \beta_0 \ln (\rho m^2(0)/\Lambda^2)$} \cite{Cornwall:1981zr,Aguilar:2006gr,Binosi:2009qm} as can be seen  on the right panel of  Fig. \ref{aperb}.
As shown in Fig.~\ref{alpha}, the coupling constant in the perturbative and non-perturbative approaches  are close in size for reasonable values of the parameters from very low $Q^2$ onward ( $Q^2 > 0.1$ GeV$^2$). This result supports the perturbative approach used up to now in model calculations, since it shows, that despite the vicinity of the Landau pole to the hadronic scale, the perturbative expansion is quite convergent and agrees with the non-perturbative results for a wide range of parameters.

In Ref.~\cite{Courtoy:2011mf} the  perturbative evolution  approach is justified by comparing  it to the non-perturbative momentum dependence as determined by the phenomenon of the freezing of the coupling constant, and to analyze the consequences of introducing an  effective gluon mass. 

\section{Non-perturbative QCD  and the Hadron Scale}
\label{sect:NP_scale}
\vspace{.5cm}

The perturbative and non-perturbative approaches can be inferred from the point of view of  hadronic models.   
 We  use, as an example, the original bag model, in its most naive description, consisting of a cavity of perturbative vacuum surrounded by non-perturbative vacuum.
The bag model is designed to describe fundamentally static properties, but in QCD all matrix elements must have a scale associated to them as a result of the RGE  of the theory.  A fundamental step in the development of the use of hadron models for the description of  properties at high momentum scales was the assertion that all calculations done in a model should have a  RGE scale associated to it \cite{Jaffe:1980ti}. The momentum distribution inside the hadron is only related to the hadronic scale and not to the momentum governing the RGE. Thus a model calculation only gives a boundary condition for the RG evolution as can be seen for example in the LO evolution equation for the moments of the valence quark distribution 
\begin{equation}
\langle q_v(Q^2)\rangle_n = \langle q_v(\mu_0^2)\rangle_n \left(\frac{\alpha_s{(Q^2)}}{\alpha_s{(\mu_0^2)}}\right)^{d^n_{NS}},
\label{moments}
\end{equation}
where $d^n_{NS}$ are the anomalous dimensions of the Non Singlet distributions.  Inside the bag, the dynamics  described by the model is unaffected by the evolution procedure, and the model provides only the expectation value,  $\langle q_v(\mu_0^2)\rangle_n$, which is associated with the hadronic scale. 
The latter is related to the maximum wavelength at which the structure begins to be unveiled.
This explanation goes over to non-perturbative evolution. The non-perturbative solution of the Dyson--Schwinger equations results in the appearence of an infrared cut-off in the form of a gluon mass which determines the finiteness of the coupling constant in the infrared. The crucial statement is that the gluon mass does not affect the dynamics inside the bag, where perturbative physics is operative and therefore our gluons inside will behave as massless. However, this mass will affect the evolution as we have seen in the case of  the coupling constant. The generalization of the coupling constant results to the structure function imply that the LO evolution Eq.~(\ref{moments})  simply changes by incorporating  the non-perturbative coupling constant  evolution Eq.~(\ref{alphalog}). 
%

%
%
 %
The non-perturbative results, using the same parameters as before, are quite close to those of the perturbative scheme and therefore we are confident that the latter is a very good approximate description.  We note however, that the corresponding hadronic scale, for the sets of parameters chosen, turns out to be slightly smaller than in the perturbative case ($\mu_0^2 \sim 0.1$ GeV$^ 2$), even  for small gluon mass $m_0 \sim 0.3$ GeV and small $\rho \sim 1$.    One could reach a pure valence scenario at higher $Q^2$ by forcing the parameters but at the price of generating a singularity in the coupling constant in the infrared associated with the specific logarithmic form of the parametrization. We feel that this strong parametrization dependence and the singularity are non physical since the fineteness of the coupling constant in the infrared is a wishful outcome of the non-perturbative analysis. In this sense, the non-perturbative approach  seems to favor a scenario where, at the hadronic scale, we have not only valence quarks but also gluons and sea quarks \cite{Scopetta:1997wk,Scopetta:1998sg}. We mean by this statement that to get a scenario with only valence quarks we are forced to very low gluon masses and very small values $\rho$, while a non trivial scenario allows more freedom in the choice of parameters.

\section{The Hadronic Scale from Perturbative Approaches}
\label{sect:soft_evo}
\vspace{.5cm}

Although the perturbative stage of a hard collision is distinct from the non-perturbative regime characterizing the hadron structure, early experimental observations suggest that, in specific kinematical regimes, both {\it the perturbative and non-perturbative stages arise almost ubiquitously}, in the sense that the non-perturbative description follows the perturbative one. In this Section, we discuss an example of perturbative approach in which the transition from perturbative to non-perturbative QCD is clearly identified.

There exists, in DIS processes, a dual description between low-energy and high-energy behavior of a same observable, {\it i.e.} the unpolarized structure functions. Bloom and Gilman observed a connection between the structure function $\nu W_2(\nu, Q^2)$ in the nucleon resonance region and that in the deep inelastic continuum~\cite{Bloom:1970xb,Bloom:1971ye}. The resonance structure function was found to be equivalent  to the deep inelastic one, when averaged over the same range in the scaling variable.  This concept is known as {\it parton-hadron duality}: the resonances are not a separate entity but are an intrinsic part of the scaling behavior of $\nu W_2$.
The meaning of duality is more intriguing when the equality between resonances and scaling happens at a same scale. 
It can be understood as {\it a natural continuation of  the perturbative to the non-perturbative representation}.

Bloom--Gilman duality implies a one-to-one correspondence between the behavior of the structure function, $F_2$, for unpolarized electron proton scattering in the resonance region, and in the perturbative QCD regulated scaling region. 
In DIS, the relevant kinematical variables are the Bjorken scaling variable, $x=Q^2/2M\nu$ with $M$ being the proton mass and $\nu$ 
the energy transfer in the lab system, the four-momentum transfer, $Q^2$, and the invariant mass for the proton, $P$, for the virtual photon, $q$, and for the system, 
%
$W^2=(P+q)^2= Q^2\left(1-x\right)/x+M^2$.
For large values of Bjorken $x  \geq 0.5$, and $Q^2$ in the multi-GeV$^2$ region, the cross section is dominated by resonance formation, {\it i.e.} $W^2 \leq 5$ GeV$^2$.
While it is impossible to reconstruct the detailed structure of the proton resonances, these remarkably follow the pQCD predictions when averaged over  the resonance region.

To answer the question of the nature of a dual description, an option is to focus on purely perturbative analysis from perturbative QCD evolution. Although Bloom--Gilman duality has been known for years, quantitative analyses could be attempted only more recently, having at disposal the extensive, high precision data from Jefferson Lab~\cite{Melnitchouk:2005zr,Liang:2004tj}. Perturbative QCD-based studies~\cite{Liuti:2011rw,Bianchi:2003hi,Liuti:2001qk}, have been presented that include  higher-twist contributions or, more generally, the evidence for non-perturbative inserts, which are required to achieve a fully quantitative fit of PDFs, especially at large-$x$. In Ref.~\cite{Courtoy:2013qca}, we discuss the Bloom--Gilman duality from a purely pertubative point of view, by analyzing the scaling behavior of the resonances at the same low-$Q^2$, high-$x$ values as the $F_2$ data from JLab. Our study leads to an analysis of the role of the running coupling constant in the infrared region in tuning the experimental data. 

A quantitative definition of  \emph{global duality} is accomplished by comparing limited intervals  defined according to the experimental data. 
Hence, we analyze the scaling results as a theoretical counterpart, or an output of perturbative QCD, in the same kinematical intervals and at the same scale $Q^2$ as the data for $F_2$. It is easily realized that the ratio,
\begin{eqnarray}
R^{\mbox{\tiny exp/th}}(Q^2)&\equiv&\frac{
\int_{x_{\mbox {\tiny min}}}^{x_{\mbox {\tiny max}}} dx\,
F_2^{\mbox {\tiny exp}} (x, Q^2)
}
{\int_{x_{\mbox{\tiny min}}}^{x_{\mbox{\tiny max}}} dx\,
F_2^{\mbox {\tiny th}} (x, Q^2)
}=1\quad,
\label{eq:ratio_1}
\end{eqnarray}
 if duality is fulfilled.\footnote{In the analysis of Ref.~\cite{Courtoy:2013qca}, we use, for $F_2^{\mbox {\tiny exp}} $, the data from JLab (Hall C, E94110)~\cite{Liang:2004tj} reanalyzed (binning in $Q^2$ and $x$) as explained in~\cite{Monaghan:2012et} as well as the SLAC data~\cite{Whitlow:1991uw}.}

Duality is violated (the ratio~(\ref{eq:ratio_1}) is not $1$) when considering the fully perturbative expression, and is still violated after corrections by the target mass terms.
One possible explanation for the apparent violation of duality is the lack of accuracy in the Parton Distribution Functions (PDF) parametrizations at large-$x$.\footnote{In our analysis, we  use the MSTW08 set at NLO as initial parametrization~\cite{Martin:2009iq}. We have checked that there were no significant discrepancies when using other sets. } Therefore, the behavior of the nucleon structure functions in the 
resonance region needs to be addressed in detail  
in order to be able to discuss 
theoretical predictions in the limit $x \rightarrow 1$. In such a limit, terms containing powers of 
$\ln (1-z)$, $z$ being the longitudinal 
variable in the evolution equations, that are present in 
the Wilson coefficient functions $B_{\mbox{\tiny NS}}^q(z)$ become large and have to be resummed, {\it i.e.} Large-$x$ Resummation (LxR).
Resummation was first introduced by  
linking this issue to the definition of the correct kinematical variable that determines the 
phase space for  real gluon emission
at large $x$. This was found to be $\widetilde{W}^2 = Q^2(1-z)/z$, 
instead of $Q^2$~\cite{Amati:1980ch}.
As a result, the argument of the strong coupling constant becomes $z$-dependent~\cite{Roberts:1999gb},
\begin{equation} 
\alpha_s(Q^2) \rightarrow \alpha_s\left(Q^2 \frac{(1-z)}{z}\right)\quad.
\end{equation}

In this procedure, however, an ambiguity is introduced, related to the need of continuing 
the value of $\alpha_s$  
for low values of its argument, {\it i.e.} for $z \rightarrow 1$. 
In Ref.~\cite{Courtoy:2013qca}, we have reinterpretated  $\alpha_s$ for values of the scale in the infrared region. 
To do so, we investigated the effect induced by changing the argument of $\alpha_s$ on the behavior of the $\ln(1-z)$-terms in the convolution with the coefficient function $B_{\mbox{\tiny NS}}$: %
\begin{eqnarray}
F_2^{NS} (x, Q^2) 
&=&x q(x,Q^2)+ \frac{\alpha_s}{4\pi} \mathlarger{\sum}_q \mathlarger{\int}_x^1 dz \, B_{\mbox{\tiny NS}}^q(z) \, \frac{x}{z}\, q\left(\frac{x}{z},Q^2\right)\quad,
\label{eq:evo}
\end{eqnarray}
We resum those terms as
\begin{eqnarray}
\ln(1-z)&=&\frac{1}{\alpha_{s, \mbox{\tiny LO}}(Q^2) }\int^{Q^2} d\ln Q^2\, \left[\alpha_{s, \mbox{\tiny LO}}(Q^2 (1-z)) -\alpha_{s, \mbox{\tiny LO}}(Q^2) \right]
\equiv \ln_{\mbox{\tiny LxR}}\quad,
\end{eqnarray}
 including the complete $z$ dependence of $\alpha_{s, \mbox{\tiny LO}}(\tilde W^2)$ to all logarithms.
Using the ``resummed"  $F_2^{\mbox{\tiny theo}} $ in Eq.~(\ref{eq:ratio_1}), the ratio $R$ decreases substantially, even reaching values lower than 1. It is a consequence of the change of the argument of the running coupling constant. At fixed $Q^2$, under integration over $x<z<1$, the scale $Q^2\times(1-z)/z$ is shifted  and can reach low values, where the running of the coupling constant starts blowing up. At that stage, our analysis requires non-perturbative information.

In the light of quark-hadron duality, it is necessary to prevent the evolution from enhancing the scaling contribution over the resonances.  We define the limit from which non-perturbative effects have to be accounted for by setting a maximum value for the longitudinal momentum fraction, $z_{max}$. Two distinct regions can be studied: the ``running" behavior  in $x<z<z_{max}$ and the ``steady" behavior $z_{max}<z<1$. 
Our definition of the maximum value for the argument of the running coupling follows from the realization of duality in the resonance region. The value $z_{max}$ is reached at
\begin{eqnarray}
R^{\mbox{\tiny exp/th}}(z_{\mbox{\tiny max}}, Q^2)
&=&\frac{
 \mathlarger{\int}_{x_{\mbox {\tiny min}}}^{x_{\mbox {\tiny max}}} dx\,
F_2^{\mbox {\tiny exp}} (x, Q^2)
}
{ \mathlarger{\int}_{x_{\mbox{\tiny min}}}^{x_{\mbox{\tiny max}}} dx\,
F_2^{NS, \mbox{\tiny Resum}} (x, z_{\mbox{\tiny max}}, Q^2)
}
= \frac{I^{\mbox{\tiny exp}}}{I^{ \mbox{\tiny Resum}}}= 1\quad.
\label{eq:ratio_max}
\end{eqnarray}

\begin{figure}[h]
\centering
\includegraphics[scale= .8]{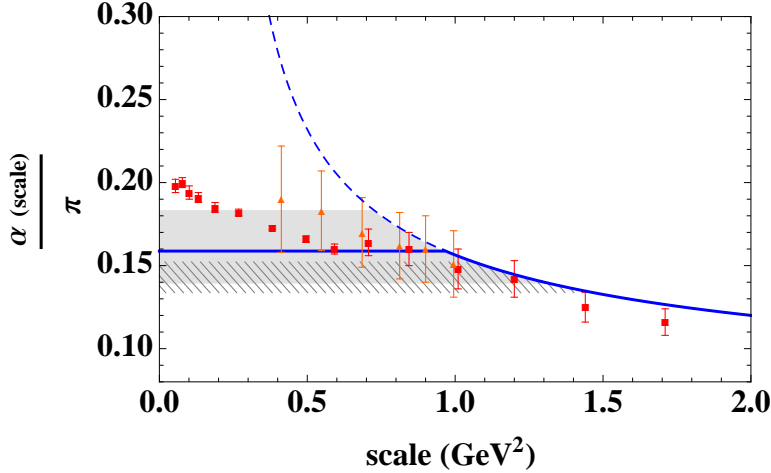}
\caption{Extraction of $\alpha_s$.  See text. 
\label{extract}}
\end{figure}
%

The direct consequence of Eq.~(\ref{eq:ratio_max}) is that duality is realized, within our assumptions, by allowing $\alpha_s$  to run from a minimal  scale only. From that minimal scale downward, the coupling constant does not run, it is frozen. This feature is illustrated on Fig.~\ref{extract}. We show the behavior of $\alpha_{s, \mbox{\tiny NLO}}$(scale) in the $\overline{\mbox{MS}}$ scheme and for the same value of $\Lambda_{\overline{\mbox{\scriptsize MS}}, \mbox{\scriptsize MSTW}}^{\mbox{\scriptsize NLO}}=0.402$ GeV used throughout our analysis. 
The theoretical errorband correspond to the extreme values of
\begin{equation}
\alpha_{s, \mbox{\tiny NLO}}\left(Q_i^2\frac{(1-z_{max, i})}{z_{max, i }}\right) \qquad,
\end{equation}
$i$ corresponds to the data points.
The determination of the transition scale $Q_0^2$ is probably the main result of our analysis. 
Of course, we expect the transition from non-perturbative to perturbative to occur at one unique scale. The discrepancy between the 10 values we have obtained has to be understood as the resulting error propagation. The grey area represents the approximate frozen value of the coupling constant,
\begin{equation}
0.13\leq \frac{\alpha_{s, \mbox{\tiny NLO}} (\mbox{scale}\to 0 \mbox{GeV}^2)}{\pi} \leq 0.18 \quad.
\end{equation}
The solid blue curve represents the (mean value of the) coupling constant obtained from our analysis using inclusive electron scattering data at large $x$.  The blue dashed curve represents the exact NLO solution for the running coupling constant in $\overline{\mbox{MS}}$ scheme.  The grey area represents the region where the freezing occurs for JLab data, while the hatched area corresponds the freezing region determined from SLAC data. This  error band represents  the theoretical uncertainty in our analysis. 

In the figure we also report  values from the extraction using polarized $eP$ scattering data in Ref.~\cite{Deur:2005cf,Deur:2008rf,alexandre}. These values represent the  first extraction of an effective coupling in the IR region that was obtained by analyzing the data  
relevant for the study of the GDH sum rule. To extract the coupling constant,  the $\overline{\mbox{MS}}$ expression of the Bjorken sum rule up to the 5th order in alpha (calculated in the $\overline{\mbox{MS}}$ scheme) was used. The red squares correspond to $\alpha_s$ extracted from Hall B CLAS EG1b, with statistical uncertainties; the  orange triangles corresponds to Hall A E94010 / CLAS EG1a
data, the uncertainty here contains both statistics and systematics.
The agreement with our analysis, which is totally independent, is impressive.
We notice, and it is probably one of the most important result of our analysis, that the transition from perturbative to non-perturbative QCD seems to occur around $1$ GeV$^2$.

At that stage, a comparison with fully non-perturbative effective charges and ``modified pQCD" is noteworthy. It is shown in Fig.~\ref{comparisons}.
The grey areas are as in Fig.~\ref{extract} with $\Lambda_{\overline{\mbox{\scriptsize MS}}, \mbox{\scriptsize MSTW}}^{\mbox{\scriptsize NLO}}=0.402$ GeV ; the dashed blue curve is the exact NLO solution with the same $\Lambda$. 
The dotted-dashed orange curve corresponds to the result of Ref.~\cite{Fischer:2003rp}, using the version $(b)$ of their
fit  with $a=b=1$. The latter analysis was performed in the MOM renormalization scheme. Though the $\beta$ function does not depend on the scheme up to 2 loops, the definition of $\Lambda$ varies from scheme to scheme. The comparison of the results is made possible using the relation,~\cite{Boucaud:1998bq}
\begin{equation}
\Lambda_{\overline{\mbox{\scriptsize MS}}}=\frac{\Lambda_{\mbox{\scriptsize MOM}}}{3.334}\quad,
\end{equation}
leading to the value of $\Lambda_{\overline{\mbox{\scriptsize MS}}}^{\mbox{\scriptsize Ref.~\cite{Fischer:2003rp}}}=(0.71/3.334) $ GeV$\sim 0.21$GeV. The value of $\alpha(0)$ is fixed to $8.915/N_c$.
The red curves are variations of the effective charge of Ref.~\cite{Cornwall:1981zr}, in Eq.~(\ref{alphalog}) with a logarithmic running for the gluon mass described by Eq.~(\ref{eq:mass})
where $(m_0^2, \rho, \Lambda)$ are parameters to be fixed. The solid red curve corresponds to the set $(m_0^2=0.3 $GeV$^2, \rho=1.7, \Lambda=0.25$GeV$)$,  the dashed red curve to $(m_0^2=0.5 $GeV$^2, \rho=2., \Lambda=0.25$GeV$)$.
This result is also obtained in the MOM scheme, the value of $\Lambda$ turns out to be similar in both Fischer {\it et al.} and Cornwall's approaches. The cyan curves correspond to two scenarios of the effective charges of Ref.~\cite{Aguilar:2009nf}. Their numerical solution is fitted by a functional form similar to Eq.~(\ref{alphalog}). The 2 sets of parameters, corresponding to $m_0=500$MeV (dashed-dotted curve) and $600$ MeV (medium dashed curve), are then driven by the shape of the numerical solution. They are plotted here with the same $\Lambda_{\mbox{\tiny MOM}}^{n_f=0}=300$ MeV as in the publication, but for $n_f=3$ for sake of comparison. Further investigation on comparison of  schemes is needed.
The short dashed green curves corresponds to Shirkov's analytic perturbative QCD to LO~\cite{Shirkov:1997wi} with $\Lambda_{\overline{\mbox{\scriptsize MS}}, \mbox{\scriptsize MSTW}}^{\mbox{\scriptsize LO}}$. The value of $\alpha(0)$ is fixed to $4\pi/\beta_0$.  Finally, the pink curve is the freezing value of Ref.~\cite{Mattingly:1992ud}. 

Notice that the freezing value for $\alpha_s(Q^2<1$GeV$^2)$ is only constrained by the integral in the resummed version of Eq.~(\ref{eq:evo}): no conclusion can be drawn on its value at $Q^2=0$GeV$^2$.   While it is not possible to conclude on the value of $\alpha_s(0)$, we notice that it is possible to find  sets of parameters for which the transition from perturbative to non-perturbative QCD occurs around $1$ GeV$^2$. 
\begin{figure}[tb]
\centering
\includegraphics[scale= 1.]{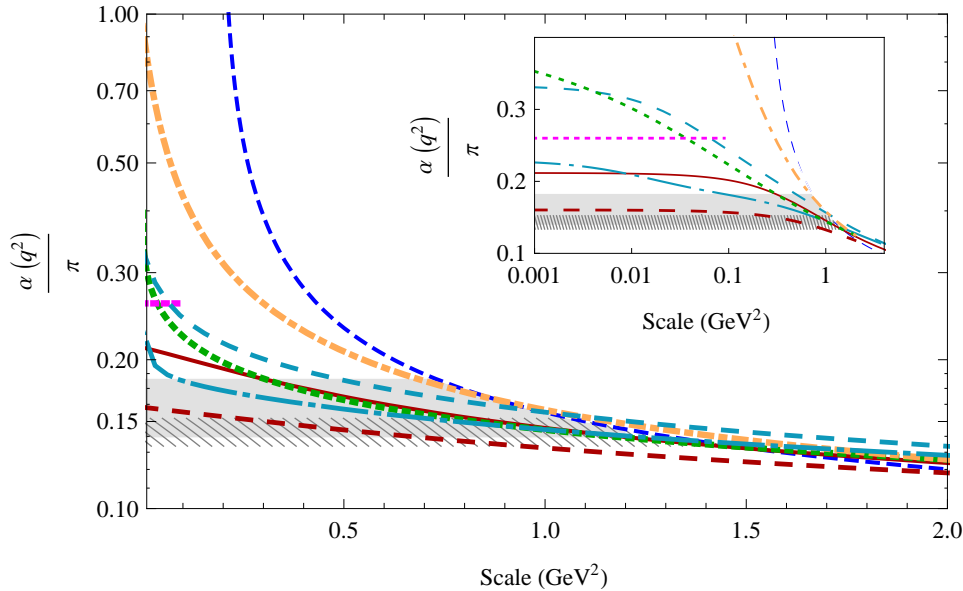}
\caption{ Comparison of the effective coupling constant from Ref.~\cite{Courtoy:2013wla}. See text. }
\label{comparisons}
\end{figure}

\section{PDFS at Large-$x$}
\label{sect:largex}
\vspace{.5cm}

As discussed in the previous Section,  the apparent violation of duality could be explained by the lack of accuracy in the Parton Distribution Functions (PDF) parametrizations at large-$x$. The remedy is obviously a better understanding and description of the large-$x$ PDFs.
The large-$x$ resummation proposed in Ref.~\cite{Courtoy:2013qca} could be implemented at the PDF level, such that, in a global fit procedure, the large-$x$ region would account for non-perturbative effects, leading to a ``cleaner" functional form for  $q(x)$.
An improved treatment of nuclear effects, especially in the resonance region,  has already been carried out by the CJ (CTEQ-JLab) collaboration~\cite{Owens:2012bv}. Independent and complementary advancements are needed as uncertainties at large-$x$ highly matter for, {\it e.g.}, search of New Physics' particles.

The impact of PDF uncertainties at large-$x$ on heavy boson production has been studied in Ref.~\cite{Brady:2011hb}, using the CJ PDF set\footnote{The new version is cited but the 2011 version was used.}.  
Hadron-hadron collisions involve at least two interacting partons,  with momentum fractions $x_1$ and $x_2$, respectively. At fixed center of mass energy $\sqrt{s}$ and boson rapidity $2\,y=\ln\left( \frac{E+p_z}{E-p_z}\right)$ where $E$ and $p_z$ are the boson energy and longitudinal momentum in the hadron center of mass frame, the parton momentum fractions are given (at leading order in the strong coupling constant) by
\begin{eqnarray}
x_{1,2} = \frac{M}{\sqrt{s}} \exp{ (\pm y)}\quad,
\end{eqnarray}
 where $M$ is the mass of the produced boson.
At low rapidities the cross sections are relatively insensitive to uncertainties in the large-$x$ PDFs. At larger rapidities, however, there is far greater sensitivity to the large-$x$ behavior, leading to $\approx 15 \%$ uncertainty in the differential cross section for $y_Z = 4$ at the LHC, and for $y_Z = 2.8$ at the Tevatron, which correspond to parton fractions of $x \approx 0.7$.
 As for the elusive $W', Z'$ bosons, it is clear now that the increasing of their mass $M$ will directly increase the relevant PDF's $x$ values, so that higher mass bosons will more readily sample the high-$x$ region where the nuclear uncertainties are more prominent. The cross sections will also decrease rapidly with increasing boson mass, so that the effects of the large-$x$ PDF uncertainties will become more significant as the mass increases. The effect on the exclusion limits from, {\it e.g.}, ATLAS~\cite{ATLAS-CONF-2012-129} is important.

\section{Tensor and Scalar Charges}
\label{sect:charges}
\vspace{.5cm}

As we have already stated in these proceedings, the hadronic structure carries important information for high energy observables. Hadronic observables are often related to the manifestation of fundamental processes at the quark level. We here discuss the structural scalar and tensor currents.

\subsection{Relation to New Physics}
\vspace{.5cm}

Non-standard electroweak couplings, studied in the decay of ultracold neutrons~\cite{Bhattacharya:2011qm}, are proposed that are related to hadronic matrix elements. Beyond the well-known weak interactions of the Standard Model, new physics'coupling could be probed in neutron $\beta$-decay. The latter are related to the isovector scalar and axial-vector hadronic matrix elements ~\cite{Weinberg:1958ut},
\begin{subequations}
\begin{eqnarray}
\left \langle p(P_p) \right | \bar{u} d \left | n(P_n)\right\rangle &=& g_S(\Delta^2)\, {\overline u}_p(P_p) u_n(P_n)\quad, \\
\left \langle p(P_p) \right | \bar{u} \sigma_{\mu\nu} d \left | n(P_n)\right\rangle &=& g_T(\Delta^2)\, {\overline u}_p(P_p) \sigma_{\mu\nu}u_n(P_n)+ \ldots\quad,
\end{eqnarray}
\end{subequations}
with $\Delta=P_n-P_p$ and the ellipsis refer to higher order terms. 
An analysis of the uncertainties in the spin-independent and spin-dependent elastic scattering cross sections of supersymmetric dark matter particles on protons and neutrons~\cite{Ellis:2008hf} concludes that the largest single uncertainty comes from the  spin-independent scattering matrix element $\left \langle N \right | \bar{q} q \left | N\right\rangle$ linked to the $\sigma_{\pi N}$ term.

The ---isoscalar and isovector--- scalar charges and the ---isovector--- tensor charge correspond to the form factors for $\Delta^2=0$, {\it i.e.}\footnote{Where we have dropped the dependence on the renormalization point.}
\begin{subequations}
\begin{eqnarray}
\langle 1\rangle_{\sigma_u-\sigma_d}&=&g_{S}(0)\quad,\\
\langle 1\rangle_{\sigma_u+\sigma_d}&=&\sigma(0)=\frac{\sigma_{\pi N}}{(m_u+m_d)/2}\quad,\label{eq:spn}\\
\langle 1\rangle_{\delta_u-\delta_d}&=&g_{T}(0)\quad.
\label{eq:isovectorTS}
\end{eqnarray}
\end{subequations}
Those matrix elements are not directly accessible through experiments, at least for $\Delta^2=0$ ; exept for the $\sigma_{\pi N}$ related to the form factor  $\sigma(2 m_{\pi}^2)$ anlyzed in Ref.~\cite{Pavan:2001wz}.

However, those charges are related to bilocal hadronic matrix elements, defining the PDFs~(\ref{eq:pdf}), through ``sum rules"~\cite{Jaffe:1992ra}.
For each quark flavor, the scalar and axial charges are related to the following forward matrix elements, respectively,
\begin{eqnarray}
\langle 1\rangle_{\sigma_q}(Q^2)&\equiv& \int_0^1 dx \left[ e^q(x,Q^2) +e^{\overline q} (x,Q^2)\right] \quad,\nonumber\\
\Rightarrow  M \,e^q(x, Q^2)&=& \int\frac{d\xi^-}{4\pi} e^{i x P^+\xi^-}\left\langle PS\right| {\overline q}(0)\, q(\xi)\left |PS\right \rangle\left |_{\xi^+=\vec{\xi}=0}\right. \quad;
\label{eq:scalarSR}
\\
\nonumber\\
\nonumber\\
\langle 1\rangle_{\delta_q}(Q^2)&\equiv& \delta q (Q^2) = \int_0^1 dx \left[ h_1^q(x,Q^2) - h_1^{\overline q} (x,Q^2)\right] \quad,\nonumber\\
\Rightarrow h_1^q(x, Q^2)&=& \int\frac{d\xi^-}{4\pi} e^{i x P^+\xi^-}\left\langle PS_{\perp}\right| {\overline q}(0)\, i\sigma^{\perp +}\gamma_5q(\xi)\left |PS_{\perp}\right \rangle\left |_{\xi^+=\vec{\xi}=0}\right.\quad,
\label{eq:transversitySR}
\end{eqnarray}
where $Q^2$ is the renormalization scale.
\\

Notice that, contrary to the vector $g_V$ or axial charge $g_A$, there  is no proper sum rule associated to the tensor ``charge" (the name in itself is inaccurate).
There are  non-vanishing anomalous dimension associated to it and the tensor charge therefore evolves with
the hard scale $Q^2$~\cite{Jaffe:1992ra}. 
The dependence on the renormalization scale is given by~\cite{Kodaira:1979ib,Artru:1989zv}, to LO,
\begin{eqnarray}
\delta q(Q^2)&=&\left[ \frac{\alpha_s(Q^2)}{\alpha_s(Q_0^2)}\right]^{-4/27} \delta q(Q_0^2)
\label{eq:evoLO}
\end{eqnarray}
where the exponent comes from, for $n_f=3$, $-2\Delta_T \gamma_{qq}^{(0)}(1)/\beta_0$ with the anomalous dimensions $\Delta_T \gamma_{qq}^{(0)}(n)=\frac{4}{3}\left(\frac{3}{2}-2[\psi(n+1)+\gamma_E]\right)$, with $\psi(n)=d \ln \Gamma(n)/dn$.

the tensor charge has been calculated on
the lattice~\cite{Gockeler:2006zu,Green:2012ud,Bhattacharya:2011qm} and in various
models~\cite{Cloet:2007em,Wakamatsu:2007nc,He:1995gz,Pasquini:2006iv,Gamberg:2001qc},
and it turns out not to be small. 
On the other hand, the $\sigma_{\pi N}$ is renormalization point invariant. The sum rule is however related to a singularity and is not of practical use.
 While the axial charge is a charge-even
operator, from Eq.~(\ref{eq:transversitySR}), it is evident that the tensor
charge is odd under charge conjugation and, therefore, it does not receive
contributions from $q\bar{q}$ pairs in the sea and is
dominated by valence contributions.

\subsection{Determination of Tensor and Scalar Charges through PDFs}
\vspace{.5cm}

The main experimental access to the tensor and scalar charges is provided thanks to the sum rules Eqs.~(\ref{eq:scalarSR}, \ref{eq:transversitySR}). The knowledge on the PDFs $h_1(x)$ and $e(x)$, up to theoretical limitations, will bring some light on the values of $g_{S/T}$.

Let's start with the transversity PDF.
A comprehensive review of the properties of the transversity distribution
function can be found in Ref.~\cite{Barone:2001sp}.
Transversity $h_1$, as leading-twist collinear PDF, enjoys the same status as
$q$ and $g_1$~\cite{Ralston:1979ys,Jaffe:1992ra}.
The distribution of transversely polarized quarks $q^{\uparrow}$
 in a transversely
polarized nucleon $p^{\uparrow}$ (integrated over transverse momentum) can be written as
\begin{equation}
f_{q^{\uparrow}/p^{\uparrow}}(x) = q(x) + {\bf S}\cdot{\bf S}_q \, h_1^q(x) \quad ,
\label{eq:radici-h1dens}
\end{equation}
in which $\bf{S}$ is the nucleon spin and $\bf{S}_q$ the quark spin.
Therefore, transversity can be interpreted as the difference between the
probability\footnote{The probabilistic interpretation is valid in the light-cone gauge.} of finding 
a parton (with flavor $q$ and momentum fraction $x$) with
transverse spin parallel and  anti-parallel to that of the transversely polarized nucleon.
The only first principle based property on the transversity distribution is the Soffer inequality. Because a probability must be positive, we get the
important
Soffer bound~\cite{Soffer:1995ww},
\begin{equation} 2|h_1^q(x,Q^2)| \leq q(x,Q^2) + g_1^q(x,Q^2) \; ,
\label{eq:radici-soffer}
\end{equation}
which is true at all $Q^2$~\cite{Bourrely:1998bx,Vogelsang:1997ak}. An analogous relation holds for antiquark distributions.

In
spin-$\textstyle{\frac{1}{2}}$ hadrons there is no gluonic function analogous
to transversity. The most important
consequence is that $h_1^q$ for a quark with flavor $q$ does not mix with
gluons in its evolution and it behaves as a non-singlet quantity; this has
been verified up to NLO, where chiral-odd evolution kernels have been studied so
far~\cite{Hayashigaki:1997dn,Kumano:1997qp,Vogelsang:1997ak}.

\begin{figure}[h]
\begin{center}
	\includegraphics[width=7cm]	{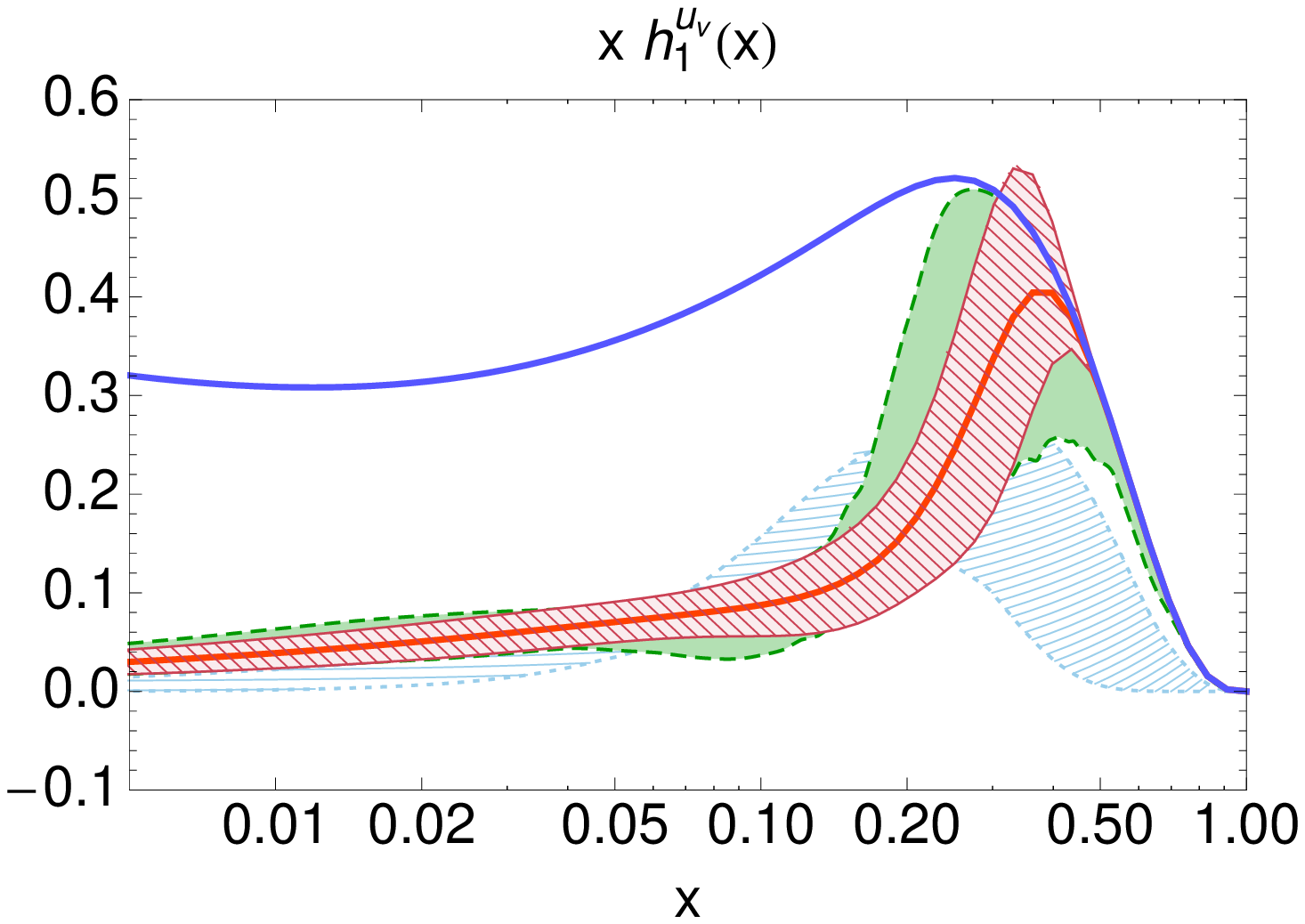}
	\includegraphics[width=7cm]	{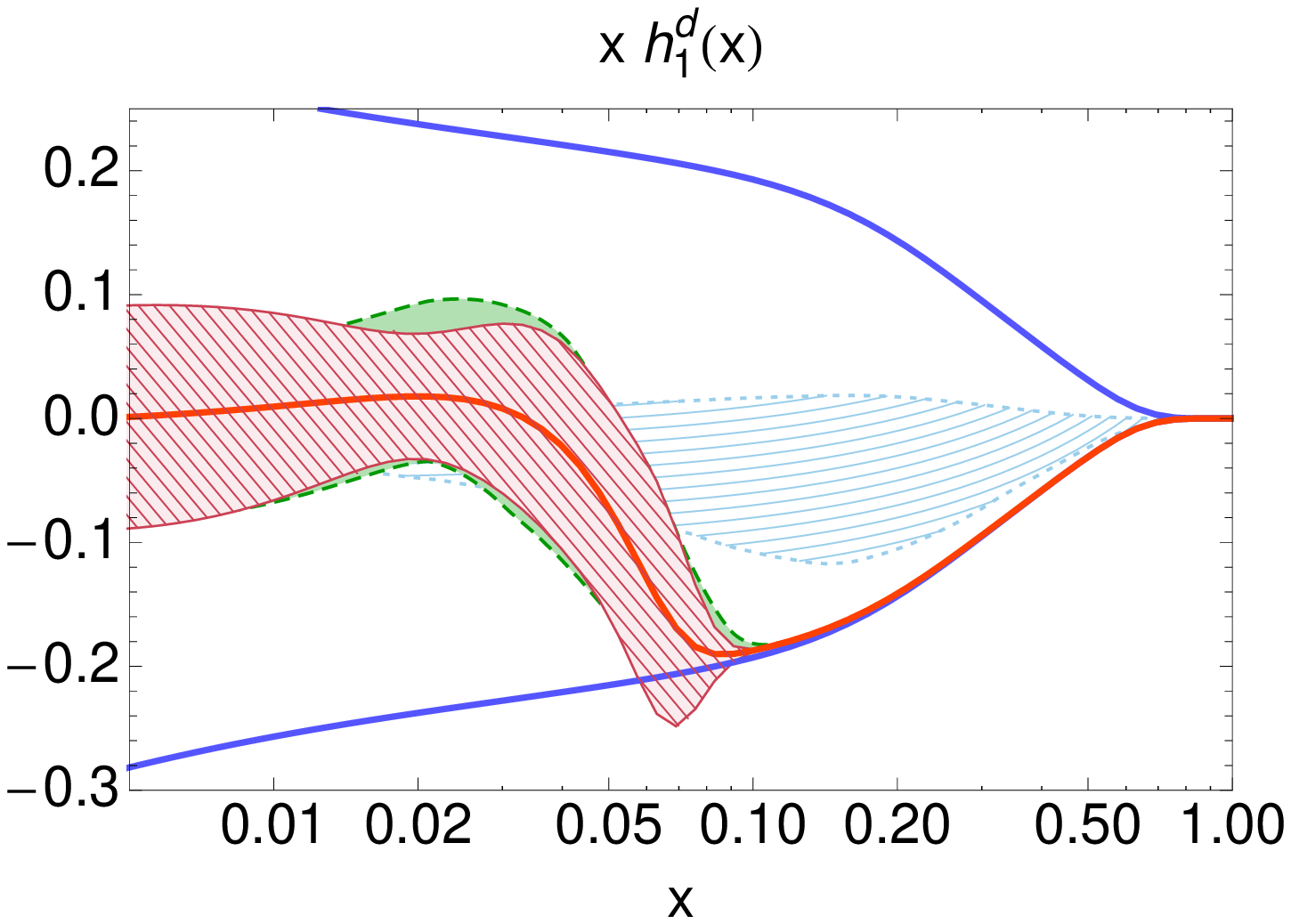}
	\includegraphics[width=7cm]	{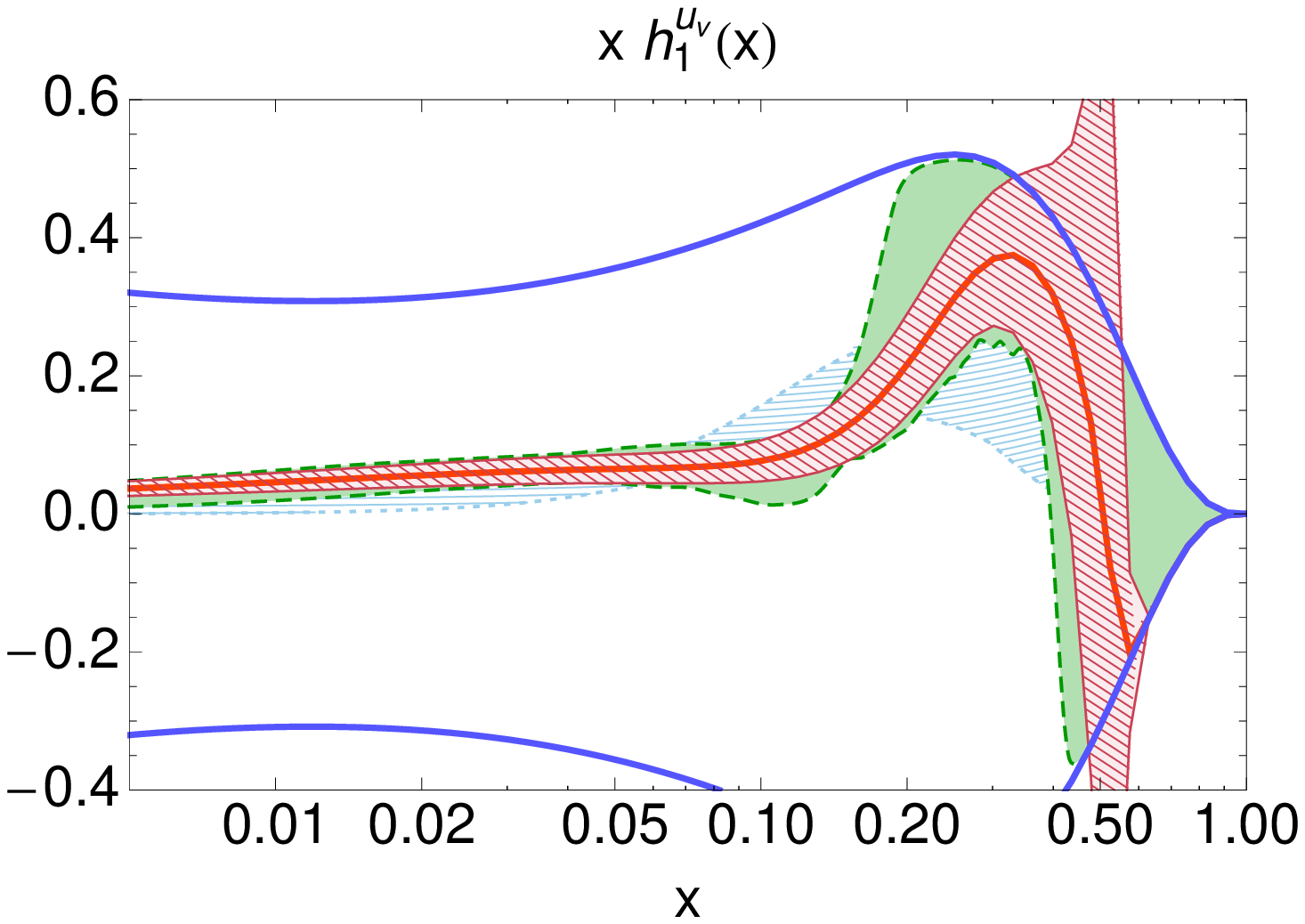}
	\includegraphics[width=7cm]	{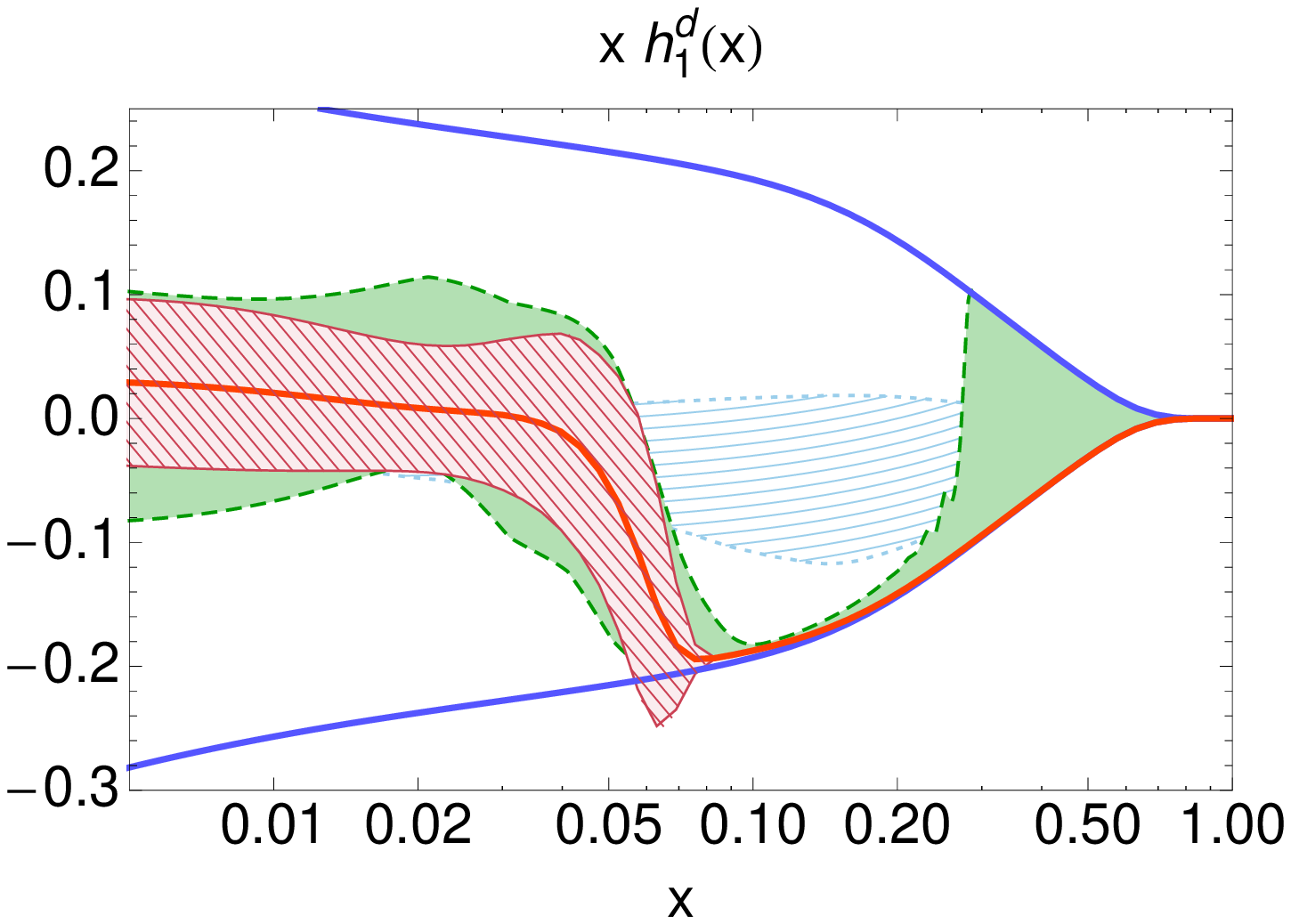}
\end{center}
\caption{Fit of valence collinear transversities at $Q^2=2.4$GeV$^2$. For $u$ and $d$ valence distributions, respectively, left and right columns. The two upper plots   correspond to the fits using a flexible parametrization, while the two lower plots correspond to a fit using an extra-flexible parameterization, see Ref.~\cite{Bacchetta:2012}. The red bands represent the standard fit within $1\sigma$ errors, while the green bands stand for a $1\sigma$  Monte Carlo based analysis ($n\times 64\%$ fit of $n$ data replica at $1\sigma$). The light blue bands correspond to the Torino13~\cite{Anselmino:2013vqa} and the full blue curves to the Soffer bound. }
\label{fig:fit_pavia}
\end{figure}

There are two complementary extractions of the transversity distributions from semi-inclusive processes: the TMD parametrization (also known as Torino fit)~\cite{Anselmino:2008jk,Anselmino:2013vqa} and the collinear extraction (also known as Pavia fit)~\cite{Bacchetta:2011ip,Bacchetta:2012}. The former is based on the TMD framework in which the chiral-odd partner of $h_1(x, k_{\perp})$ is the Collins fragmentation function ; the latter is based on a collinear framework, involving the chiral-odd dihadron fragmentation function, {\it i.e.} $H_1^{\sphericalangle}$.  Parameterizations for the dihadron FFs have been obtained independently~\cite{Courtoy:2012ry}. One of the main differences lie in that the collinear extraction does not require the use of a fitting functional form: it is a point-by-point extraction. However, for practical reasons, a statistical study of the transversity PDF has been performed as well. So far, both approaches has found compatible results in the range in $x$ where data exist. However, recent progress on TMD evolution are expected to affect the Torino fit. It is important to notice that the parameterizations are biased by the choice of  the fitting functional form. The behavior of the best-fit parametrization is largely unconstrained outside the range of data, leading to confusing results at low and large-$x$ values. This is nicely illustrated by the collinear ---Pavia--- transversity collaboration, on Fig.~\ref{fig:fit_pavia}, where 2 different functional forms, with an equally good $\chi^2/d.o.f.$, have been used.

Another independent access to the tensor charge has recently been published using exclusive processes~\cite{Goldstein:2014aja}. Through yet another ``sum rule", the tensor charge corresponds to the first Mellin moment of the chiral-odd $H_T$ generalized parton distribution.

\begin{figure}[htb]
\begin{center}
\includegraphics[height=6.5cm]{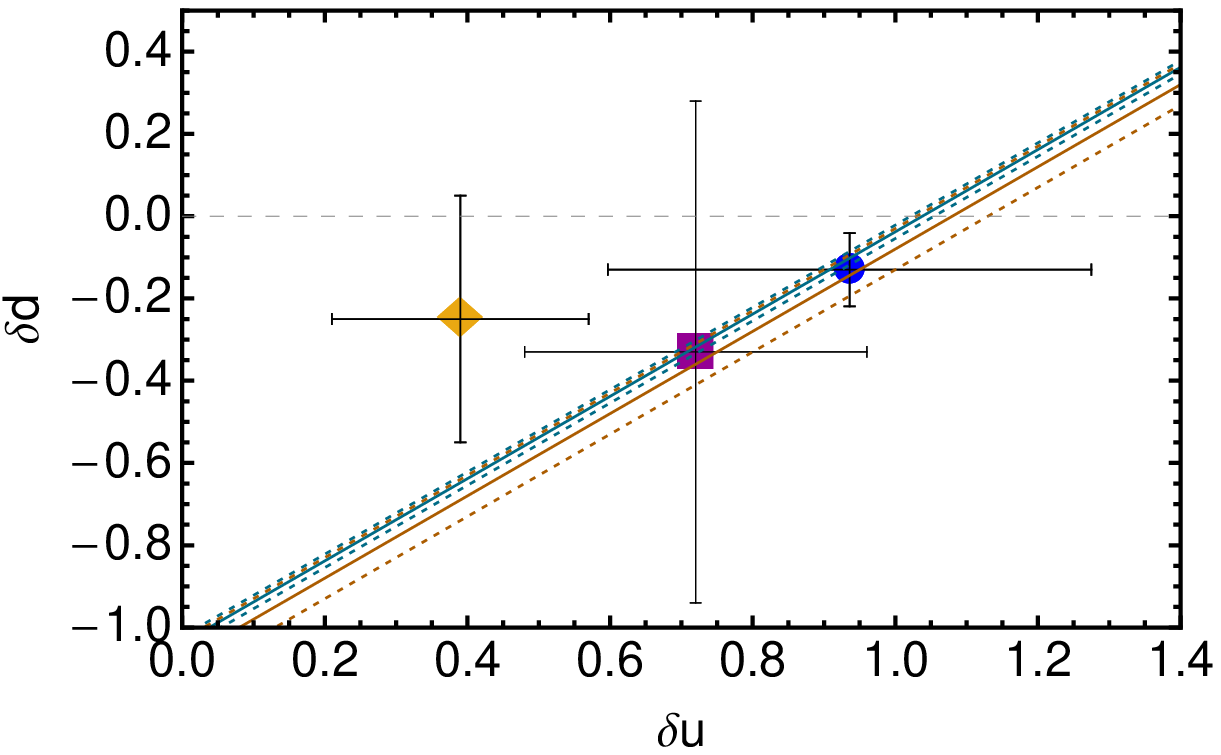}
\end{center}
\caption{
The tensor charge $\delta u$ vs. $\delta d$, computed using the transversity
distributions from TMD collaboration~\cite{Anselmino:2013vqa} (standard) (yellow diamond). The blue circle comes from the chiral-odd GPD $H_T$ sum rule, with the GPD fits of Ref.~\cite{Goldstein:2014aja}.
The purple square corresponds to the standard flexible version of the fit via DiFF~\cite{Bacchetta:2012}, see Fig.~\ref{fig:tenscharge_new} for comparison of the 2 fits' results. The cyan curve corresponds to the lattice result from Ref.~\cite{Green:2012ud} ; the brown curve to Ref.~\cite{Bhattacharya:2013ehc}. }
\label{fig:tenscharge}
\end{figure}

\begin{figure}[htb]
\begin{center}
\includegraphics[width=7.cm]{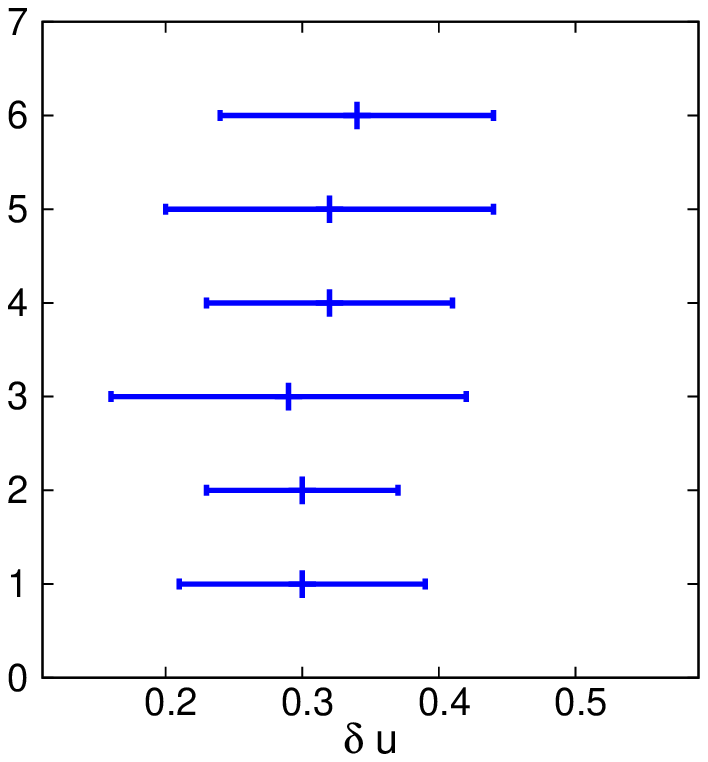}
\includegraphics[width=7.cm]{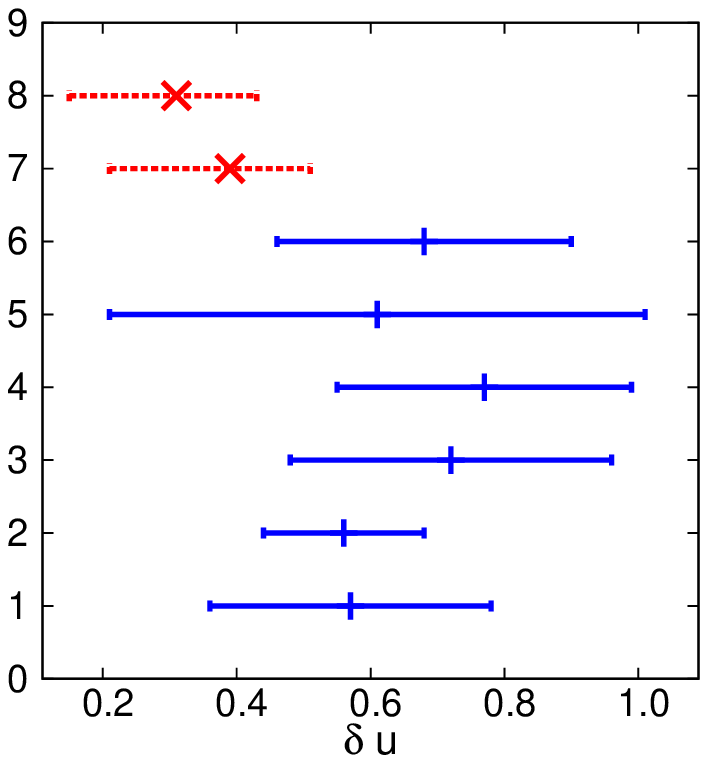}
\end{center}
\caption{{\it Left panel}: tensor charge integrated over data range for the u-quark. {\it Right panel}: tensor charge integrated on the theoretical support $x\in [0,1]$. Respectively for, from 1 to 8, standard rigid, Monte Carlo rigid, standard flexible, Monte Carlo flexible, standard extra-flexible, Monte Carlo extra-flexible of the collinear fit~\cite{Bacchetta:2012} and the fit for $A_0$ and $A_{12}$ asymmetries at Belle combined with single-hadron SIDIS of the TMD fit collaboration~\cite{Anselmino:2013vqa}.}
\label{fig:tenscharge_new}
\end{figure}

In Fig.~\ref{fig:tenscharge}, we summarize the current status on  the tensor charge for up and down quarks. The three extractions are shown at the scale used by each group ; the lattice results are shown for $Q^2=2$ GeV$^2$. In Fig.~\ref{fig:tenscharge_new}, we illustrate the strong dependence on the functional form. More data, especially in the low and large $x$ regions, are needed to constrain the fits. Hopefully, more data from PHENIX are soon to be released. Worth to mention too are the proposals at JLab ---both for CLAS12 and SoLID~\cite{ourprop,ourloi}.


As for the scalar charges, they are, though in a more elusive way, related to the twist-3 PDF $e(x)$. See Ref.~\cite{Efremov:2002qh} for a  review of the chiral-odd twist-3 PDF. QCD equations of motion allow to decompose the chiral-odd twist-3 distributions into 3 terms
\begin{eqnarray}
e^q(x)&=&e_{\mbox{\tiny{loc}}}^q(x)+e_{\mbox{\tiny{tw-3}}}^q(x)+e_{\mbox{\tiny{mass}}}^q(x)\quad.
\end{eqnarray}
The first term comes from the local operator and is related to the pion-nucleon sigma term,
\begin{eqnarray}
e_{\mbox{\tiny{loc}}}^q(x)&=&\frac{1}{2M}\, \int \frac{d\lambda}{2\pi}\, e^{i\lambda x}\langle P| \bar{\psi}_q(0)\psi_q(0)|P\rangle\quad,\nonumber\\
&=&\frac{\delta(x)}{2M}\langle P| \bar{\psi}_q(0)\psi_q(0)|P\rangle\quad;
\end{eqnarray}
the second term is a genuine twist-3 contribution, {\it i.e.} pure quark-gluon interaction term ; while the last term is related to the current quark mass.

Owing to Eq.~(\ref{eq:spn}), $\sigma_{\pi N}$ is related to $e(x=0)$, due to the delta-function singularity.
Furthermore, being a subleading contribution, this twist-3 PDF is hardly known. It is however accessible in single-~\cite{Efremov:2002ut,Gohn:2014zbz} and two-hadron~\cite{silvia_analisi} semi-inclusive DIS off unpolarized targets, through the Beam Spin Asymmetry $A_{LU}$ at CLAS. Just like in the case of transversity extraction, the chiral-odd partner of $e(x)$ is the chiral-odd dihadron fragmentation function, {\it i.e.} $H_1^{\sphericalangle}$. The analysis of the soon-to-be-released data is ongoing~\cite{CourMiraPisa}.
\\

The bounds on $g_S, g_T$, together with the bounds on beta decay couplings, $\tilde{C}_S, \tilde{C}_T$, constrain the new effective couplings, $\epsilon_S, \epsilon_T$, through the matching conditions from a quark-level effective theory  to a nucleon-level effective theory~\cite{Bhattacharya:2011qm}
\begin{eqnarray}
\tilde{C}_S&=&g_S \epsilon_S\quad,\nonumber\\
\tilde{C}_T&=&4 g_T \epsilon_T\quad,\nonumber
\end{eqnarray}
The bounds resulting from the phenomenological extractions of the tensor charge~\cite{Bacchetta:2012,Anselmino:2013vqa,Goldstein:2014aja} on the observability of new physics are still huge compared to that the lattice calculations can nowadays achieve. An analysis of the precise projection of these bounds, together with the expected precision of future hadronic structure dedicated experiments is ongoing~\cite{CourGonzLiuti}.

\section{Conclusions}
\vspace{.5cm}

Hadronic physics is the perfect framework to study the intersection between perturbative and non-perturbative QCD. The properties of hadrons reflect the features of the strong interactions and transition from chiral symmetry to confinement. 

Non-perturbative QCD provides inputs for perturbative calculation, mainly through {\it initial conditions} for the QCD evolution equations. As described in these proceedings, Parton Distribution Functions in QCD have a scale dependence, through the RGE.  Non-perturbative predictions often come from evaluation in models for the proton structure.
This will, in turn, strongly affect the PDF fits: it is called {\it procedural bias} in Ref.~\cite{JimenezDelgado:2012zx} and is a problem known in the hadronic community for years, {\it e.g.}~\cite{Traini:1997jz}. 

Besides the uncertainty related to the {\it hadronic scale} and the usual statistical errors, we have mentionned the poor knowledge on PDFs at large values of Bjorken-$x$. The uncertainty on the region $x\to 1$ \cite{Brady:2011hb} will affect the exclusion limit for heavy boson like $Z', W'$.

Finally we have mentionned hadronic matrix elements related to New Physics observables. 

The understanding of parton distributions at low energy have important repercussions for the high energy physics. 
This is why this plethora of distributions and new information about the hadron structure require to be handled carefully via complementary theoretical approaches.
Therefore, in that sense, QCD must be considered as a whole, from the infrared to the ultraviolet region.

\ack

I am grateful to Ted C. Rogers for priceless discussions about the impact of non-perturbative physics on perturbative QCD. I also thank  Simonetta Liuti and Vicente Vento, for their advices and support ; Mart\'in Gonz\'alez Alonso for the explanations about his recent works ; Alexei Prokudin and Stefano Melis for sharing the Torino13 transversity.
This work was funded by the Belgian Fund F.R.S.-FNRS via the contract of Charg\'ee de recherches.

\section*{References}

\bibliography{auro}
\end{document}